\documentclass[11pt]{article}
\usepackage{moriond,epsfig}

\def\be{\begin{equation}}
\def\ee{\end{equation}}
\def\bea{\begin{eqnarray}}
\def\eea{\end{eqnarray}}

\def\vub{$\mathrm{|V_{ub}| \,\,}$}
\def\vcb{$\mathrm{|V_{cb}| \,\,}$}
\def\dstln{$\mathrm{D^{\star}} \it{l} \nu \,\,$}
\def\gev{\mathrm{\,GeV}}
\def\btou{$b \to u \,\,$}
\def\btoc{$b \to c \,\,$}
\begin{document}
\vspace*{4cm}
\title{\vub and \vcb from CLEO}

\author{ A. Bornheim \\ (representing the CLEO Collaboration  \footnote{Presented at the {\em XXXVIII. Rencontres de Moriond, Electroweak Interactions and Unified Theories}})}

\address{Caltech, Lauritsen Laboratory, 1200 E. California Blvd. ,
Pasadena CA, 91125, USA}

\maketitle\abstracts{
We report on studies of exclusive and inclusive semileptonic
$b \to u\ell\nu$ and $b \to c\ell\nu$ decays in 9.7 million
$B\bar B$ events accumulated with the CLEO detector in symmetric
$e^+e^-$ collisions produced in the Cornell Electron Storage Ring
(CESR).  Various experimental techniques, including the inference
of neutrino candidates using the hermeticity of the CLEO detector
and the study of spectral moments, are used in conjunction with
theoretical calculations to provide estimates of the CKM matrix
elements $|V_{ub}|$ and $|V_{cb}|$.}

%
%
\section{Introduction}
\begin{figure}[h!]
 \begin{center}
  \vspace*{-1.0cm}
  \hspace*{-0.5cm}
  \begin{minipage}[l]{0.5\textwidth}
   \psfig{figure=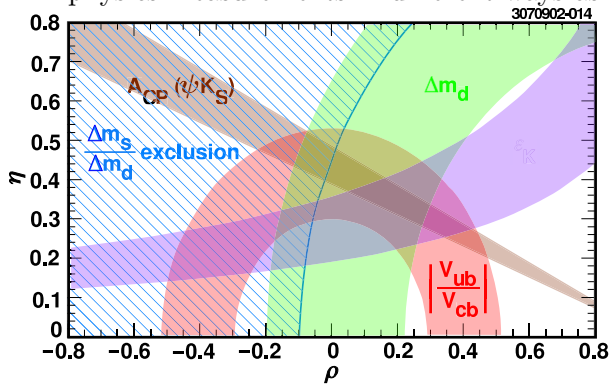,height=5.0cm}
  \end{minipage}
  \hfill
   \raisebox{0.4cm}
   {
   \begin{minipage}[r]{0.46\textwidth}
   \caption{ The $\rho$-$\eta$ plane showing the constraints on these two parameters from various measurements
    in $B$- and $K$-physics.
    The ratio of the measurements of \vub and \vcb we report on in this document provide a
    donut-shaped constraint around the origin. The measurement of $A_{CP}(\Psi /K_s)$ performed by the Belle
    and BaBar collaborations is currently still statistically limited as is the measurement of 
    $\frac{\Delta_{m_s}}{\Delta_{m_d}}$ which will be improved with the upcoming results from 
    the Tevatron run II. The measurement of $ | \epsilon_K | $ is systematically limited.
    \label{fig:rhoetaplane}}
   \end{minipage}
   }
   \vspace*{-0.2cm}
 \end{center}
\end{figure}

 One of the main goals of B-physics is to test the consistency of the standard model (SM) description
 of quark mixing \cite{ckm}.  
 In the Wolfenstein representation \cite{wolfenstein} the quark mixing matrix is described by four parameters. 
 Two of the free
 parameters, $\rho$ and $\eta$, are commonly pictured as the Unitarity Triangle. The $\rho$-$\eta$ parameter
 space is constrained by various B-physics and K-physics measurements in different ways as shown in            
 Fig.~\ref{fig:rhoetaplane}.\\  
\\
%
%
 All these measurements pose very different experimental, phenomenological and theoretical challenges.
 To measure \vub and \vcb one has to measure decay rates corresponding to the quark-level transitions
 \btou and \btoc. In practice the 
 problem is that the measurements of these rates are obscured by hadronization effects and interactions
 between initial and final state particles involved. Since no reliable method exists to date to fully calculate
 these effects one has to control them in other ways. One common approach is to use semileptonic decays
 $B \to X_ul\nu$ and $B \to X_cl\nu$ which reduce non-calculable hadronic effects considerably.   
 Furthermore one measures \vub and \vcb in many different ways in order to get a better handle 
 on the remaining non-perturbative effects. Finally one uses calculations, in particular 
 from heavy quark effective theory (HQET) \cite{hqet} and lattice QCD (LQCD) \cite{lqcd}, to link various 
 measurements together to further reduce theoretical uncertainties.  
 In the following we give an overview of the 
 \vub and \vcb measurements performed with the CLEO detector \cite{cleodet} at the CESR $e^+e^-$ collider.   
  
%
%
\section{Measurements of \vcb}
\subsection{\vcb exclusive}

\begin{figure}[h!]
\begin{center}
  \vspace*{-1.0cm}
  \begin{minipage}[r]{0.5\textwidth}
    \hspace*{1.0cm}
    \psfig{figure=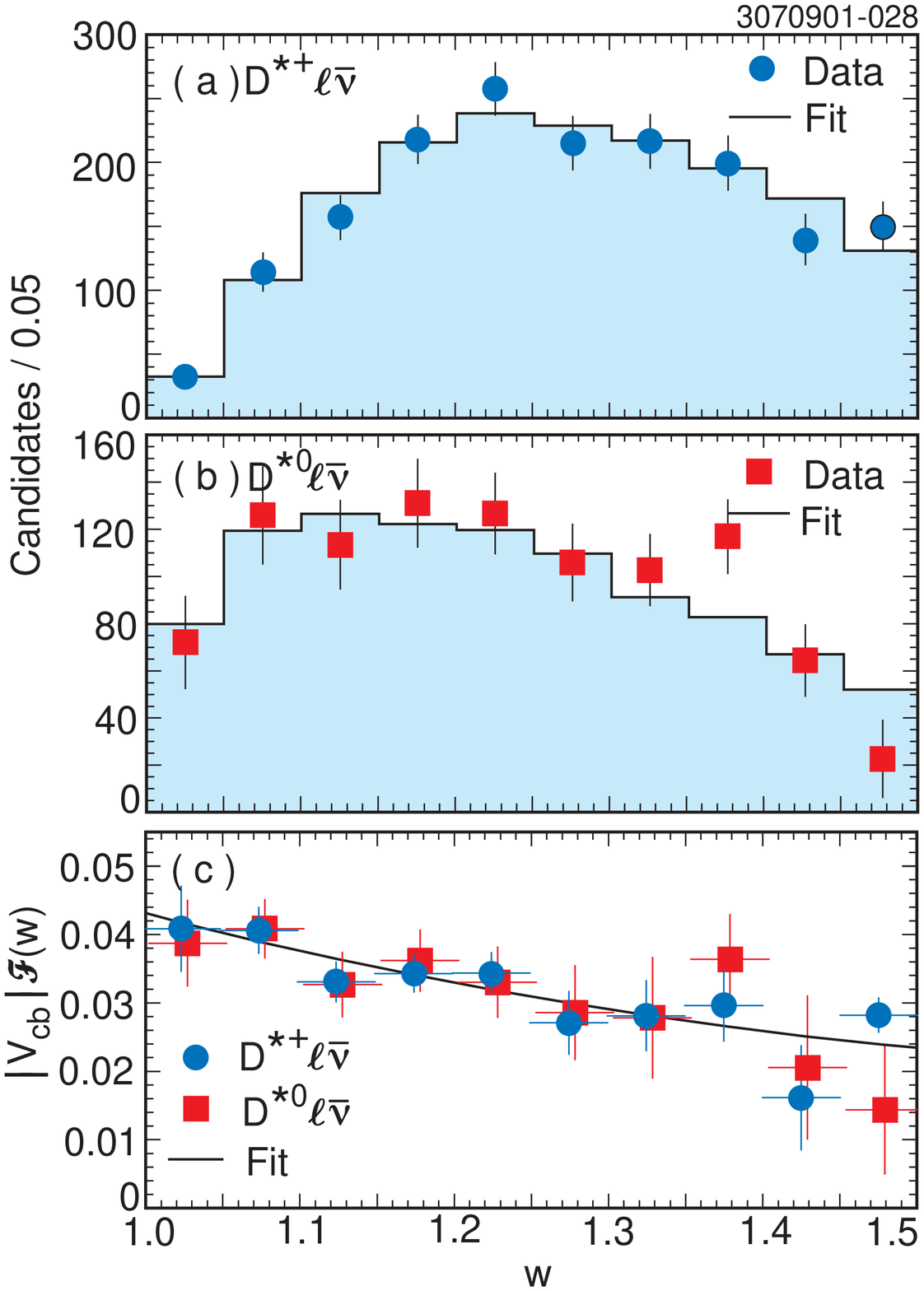,height=8.0cm}
   \vspace*{-0.5cm}
   \caption{ \dstln rate for charged (top) and neutral (center) $D^{\star}$ and \vcb ${\cal F}(w)$
            as a function of $w$ (bottom).
   \label{fig:dstarlnu}}
   \end{minipage}
  \hfill
  {
   \begin{minipage}[c]{0.44\textwidth}
     \psfig{figure=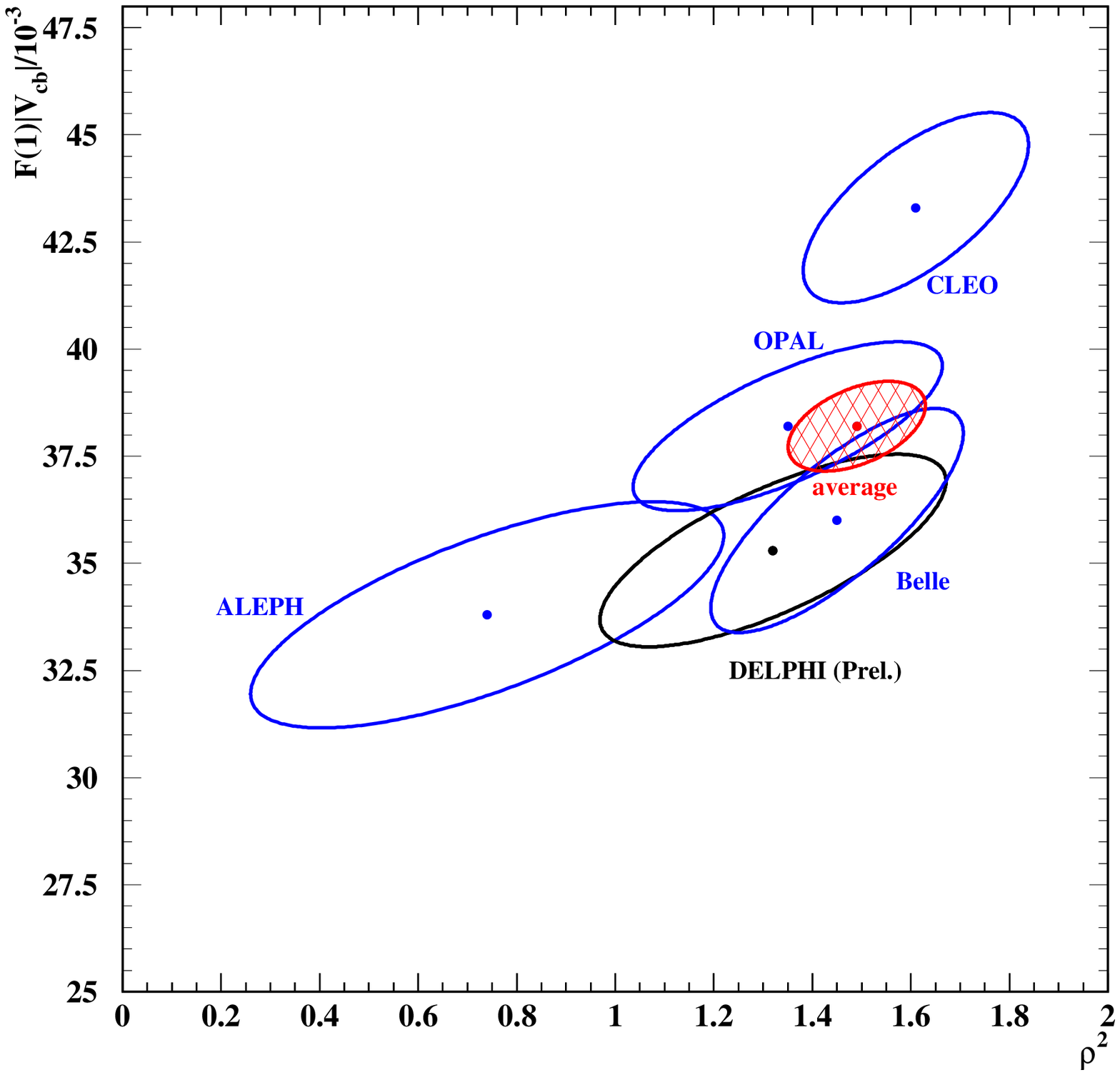,height=7.0cm}
     \vspace*{-0.7cm}
     \caption{Comparison of $|V_{cb}|$ measurements from various experiments.
       Plotted is ${\cal F}(1)|V_{cb}| \times 10^3$ versus the from-factor slope parameter $\rho^2$.  
     \label{fig:vcbcomp}}
   \end{minipage}
  }
 \end{center}
 \vspace*{-0.4cm}
\end{figure}

Using exclusive $B \to D^{\star}l\nu$ decays one can employ heavy quark symmetry relations to calculate 
strong interaction form factors that enter the decay rate.
In the framework of HQET it is useful to consider the kinematic variable 
$w = v_B \cdot v_{D^{\star}} = \frac{m^2_B+m^2_{D^{\star}}-q^2}{2m_Bm_{D^{\star}}}$, which is linearly
related to $q^2$, the mass of the virtual $W$.
The decay rate can then be written as
$\frac{d\Gamma}{d\omega}=\frac{G^2_F}{48\pi^3}|V_{cb}|^2[{\cal F}(w)]^2 {\cal K}(w)$
where ${\cal K}(w)$ is a kinematic function of the masses and the variable $w$ depends only on the \mbox{V-A} structure of the 
weak transition. ${\cal F}(w)$ is a form factor representing the strong interaction dynamics of the 
$B \to D^{\star}$ transition \cite{dstarshapeqcd}. In HQET ${\cal F}(w)$ can be calculated for $w=1$; Lattice QCD 
and QCD sum rules give similar results \cite{dstarfromlatticesumrule}. 
The shape of the form factor is less determined, QCD
dispersion relation may be used to constrain it \cite{dstarshapeqcd}.
Experimentally one measures the decay rate as a function of $w$ and extrapolates back to $w=1$ where 
${\cal F}(1)|V_{cb}|$ can then be extracted. In Fig.~\ref{fig:dstarlnu} the measured rates are shown for 
$D^{\star +}l\nu$ (top) and $D^{\star 0}l\nu$ (center). The bottom plot shows the fit to the values 
of $|V_{cb}|^2{\cal F}(w)$ derived from the corrected rates for both $D^{\star +}l\nu $ and $D^{\star 0}l\nu $.\\   
\\
From the data in the bottom plot of Fig.~\ref{fig:dstarlnu} 
we extract  ${\cal F}(1)|V_{cb}| = (4.31\pm0.13\pm0.18) \times 10^{-2} $
and the form-factor slope parameter 
$ \rho^2 = 1.61 \pm 0.09 \pm 0.21 $ with a $\chi^2$ fit. The curvature of ${\cal F}(w)$ is constrained 
as predicted from \cite{dstarshapeqcd}.
Using the charged to neutral ratio for B-mesons at the $\Upsilon (4S) \,$,  $f_{+-} = 0.521 \pm 0.012 \,$, 
to average over both modes and ${\cal F}(1) = 0.919 ^{+0.030}_{-0.035} $ from HQET we extract \cite{cleodstar} :\\
\vspace*{-0.7cm} \begin{center} $ |V_{cb}| = (4.69 \pm 0.14 \pm 0.20 \pm 0.18) \times 10^{-2} $ \end{center}
The errors are statistical, systematic and theoretical.
In Fig.~\ref{fig:vcbcomp} we compare our result with similar analysis from the LEP experiments and from BELLE.
One should take into account that there are differences between the 
analysis. The efficiency for the pion detection from the subsequent $D^{\star}$ decay
are different for the LEP- and $\Upsilon (4S)$-experiments, CLEO measures both charged an neutral $D^{\star}$, 
LEP only charged ones, and CLEO fits for background from $\bar{B} \to D^{\star}Xl\bar{\nu}$ 
while LEP uses a model. A  $2\, \sigma$-
fluctuation in the latter alone can explain the differences between LEP and CLEO \cite{vcbminireview}. This 
example illustrates very
well the importance of over constraining not only each parameter of the mixing matrix but also each 
measurement in order to get a really good understanding of all systematic aspects which limit the 
precision of the measurements. 

%
%
\subsection{\vcb inclusive}

\begin{figure}[!h]
\begin{center}
 \vspace*{-0.5cm}
 \begin{minipage}[c]{0.45\textwidth}
  \psfig{figure=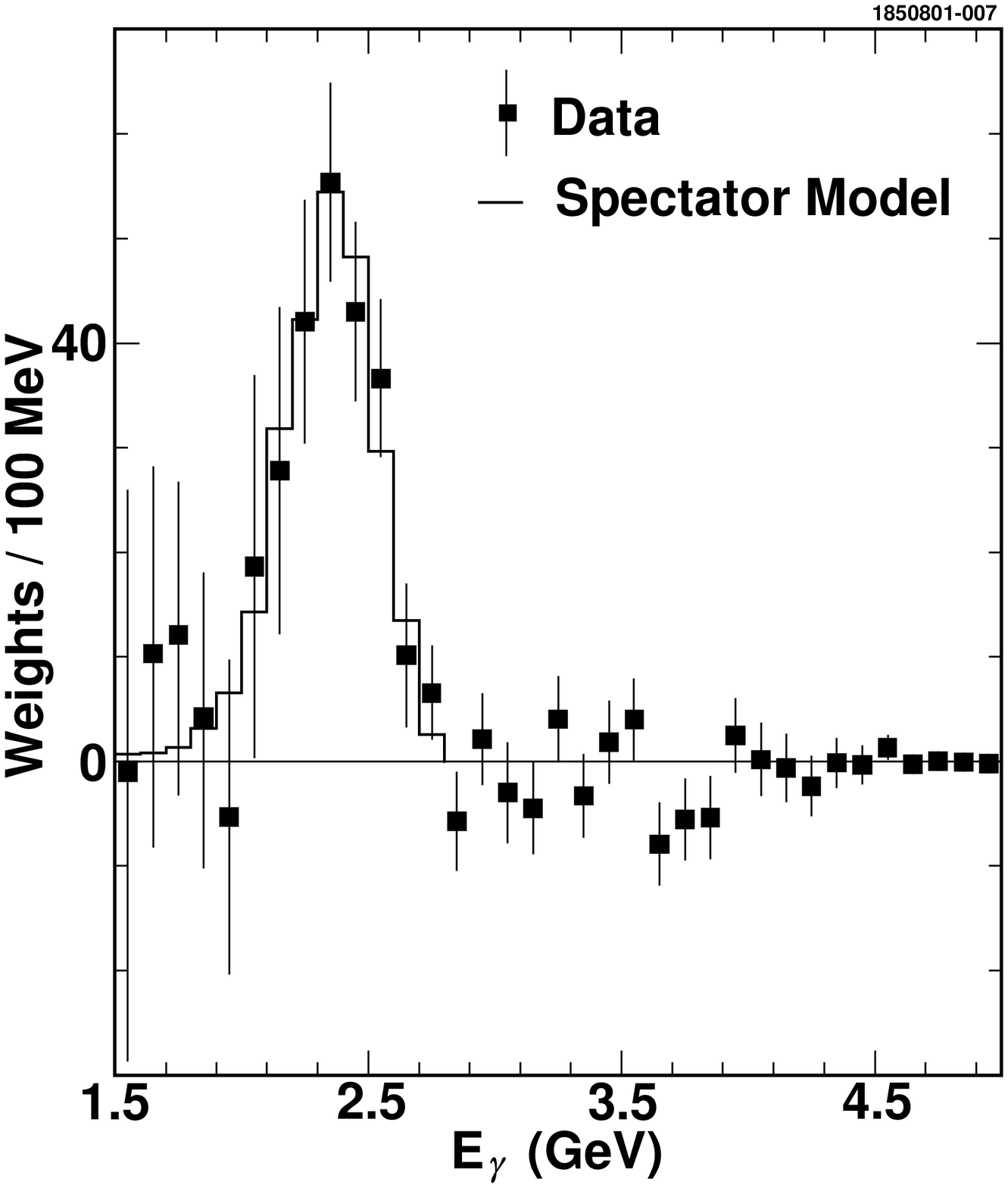,height=7.3cm}
  \vspace*{-0.4cm}
  \caption{The photon energy spectrum from inclusive $b \to s\gamma$ decays as measured 
           by CLEO.  
  \label{fig:bsgspectrum}}
 \end{minipage}
 \hfill
  {
  \begin{minipage}[c]{0.49\textwidth}
   \psfig{figure=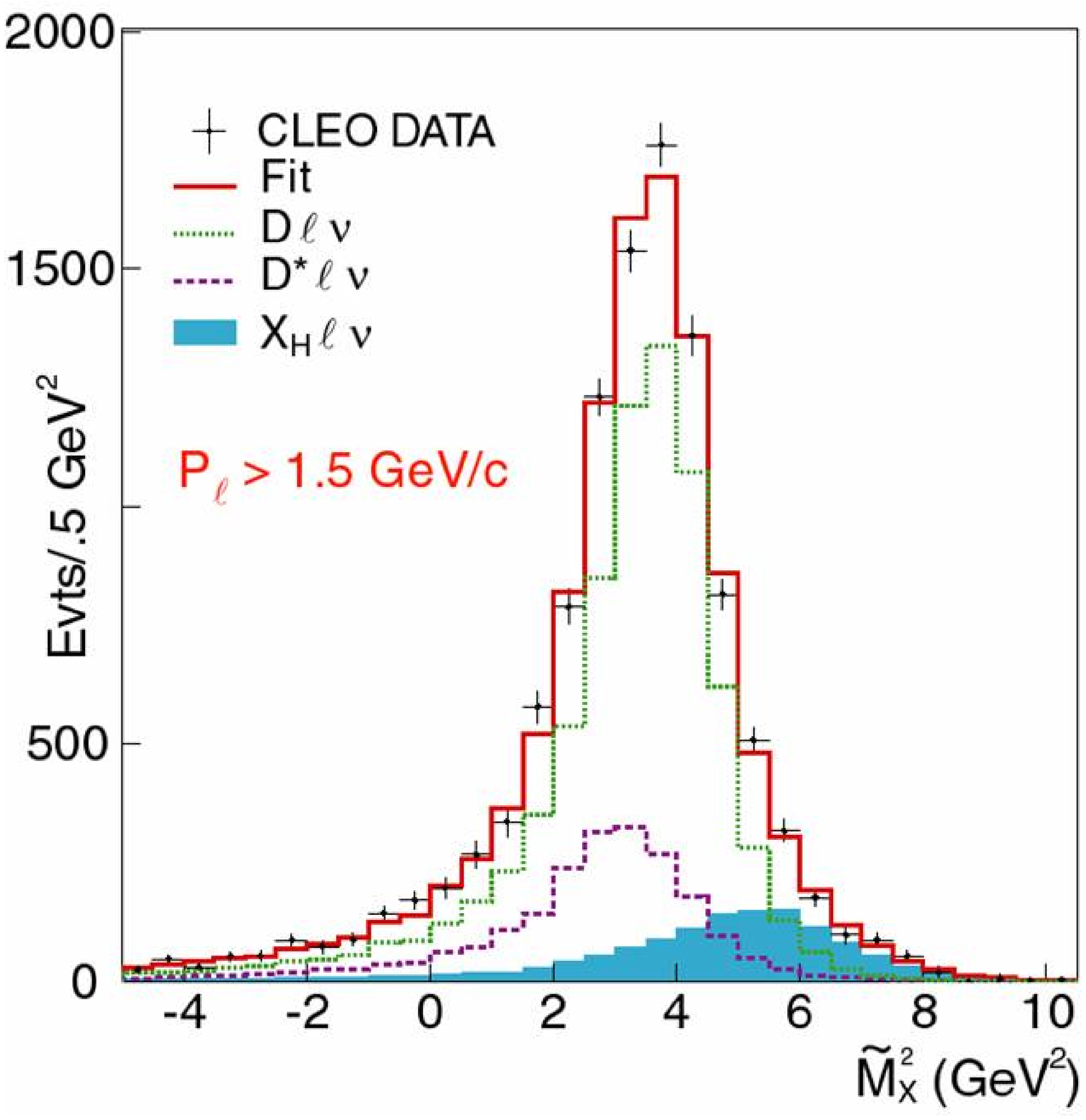,height=7.3cm}
   \vspace*{-0.4cm}
   \caption{The mass spectrum $M^2_x$ for $B \to X_c l\nu$ transitions as measured by CLEO.
   \label{fig:massmoments}}
  \end{minipage}
  }
 \end{center}
 \vspace*{-0.5cm}
\end{figure}

\begin{figure}[!h]
 \begin{center}
  \vspace{-0.8cm}
  \begin{minipage}[c]{1.0\textwidth}
    \begin{minipage}[c]{0.5\textwidth}
     \psfig{figure=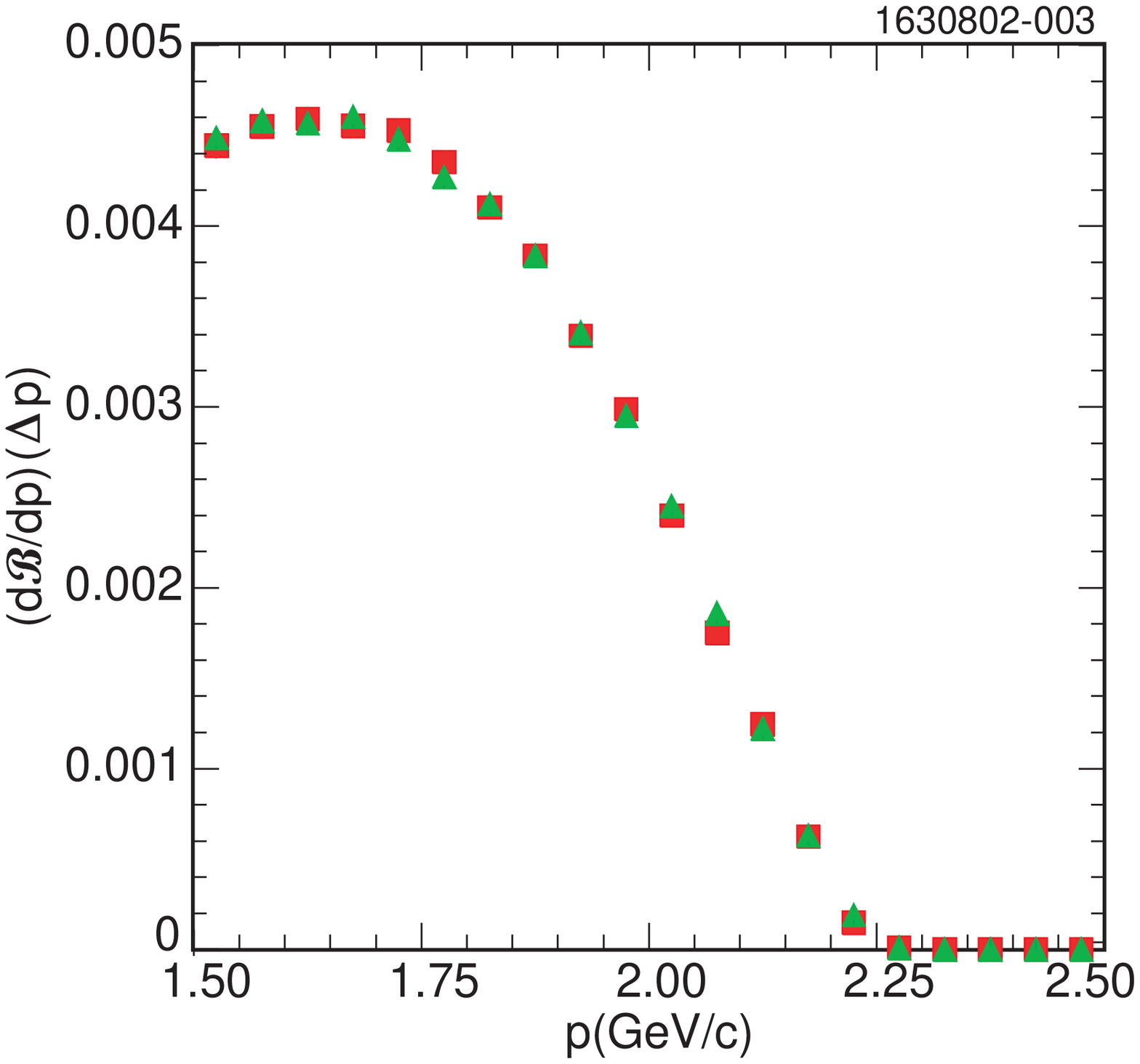,height=7.0cm}
    \end{minipage}
   \hfill
   \raisebox{2.2cm}
   {
   \begin{minipage}[r]{0.5\textwidth}
    \caption{(left) The corrected energy spectrum at the endpoint for muons (red squares) and electrons 
                    (green triangles) in the $B$-meson rest-frame from semileptonic \btoc transitions.
                    $d{\cal B}$ represents the differential semileptonic branching fraction in the bin
                    $\Delta p$, divided by the number of $B$-mesons in the sample.
            \label{fig:leptonenergy}}
   \end{minipage} 
   }
   \vspace*{-2.8cm}
  \end{minipage}
  \begin{minipage}[c]{1.0\textwidth}
   \hfill
    \begin{minipage}[l]{0.46\textwidth}
    \caption{(right) Two-dimensional representation of the \mbox{$\lambda_1$-$\bar{\Lambda}$} plane with
            experimental results and their errors on these observables shown as bands and the 
            central value and its error shown as an ellipse.
            The solid vertical band is from the
            first moment of the photon energy spectrum in inclusive $b \to s\gamma$ decays.
            The solid diagonal band is from the first moment of the mass spectrum in inclusive 
            semileptonic \btoc transitions. The constraints from the lepton energy spectrum
            is indicated by lines. All bands overlap in a single region, demonstrating the 
            self-consistency of the HQET parameter extraction. 
            \label{fig:leptonmoments}}
    \end{minipage} 
   \raisebox{1.0cm}
   {
    \begin{minipage}[c]{0.5\textwidth}
     \hspace*{-0.2cm}
     \psfig{figure=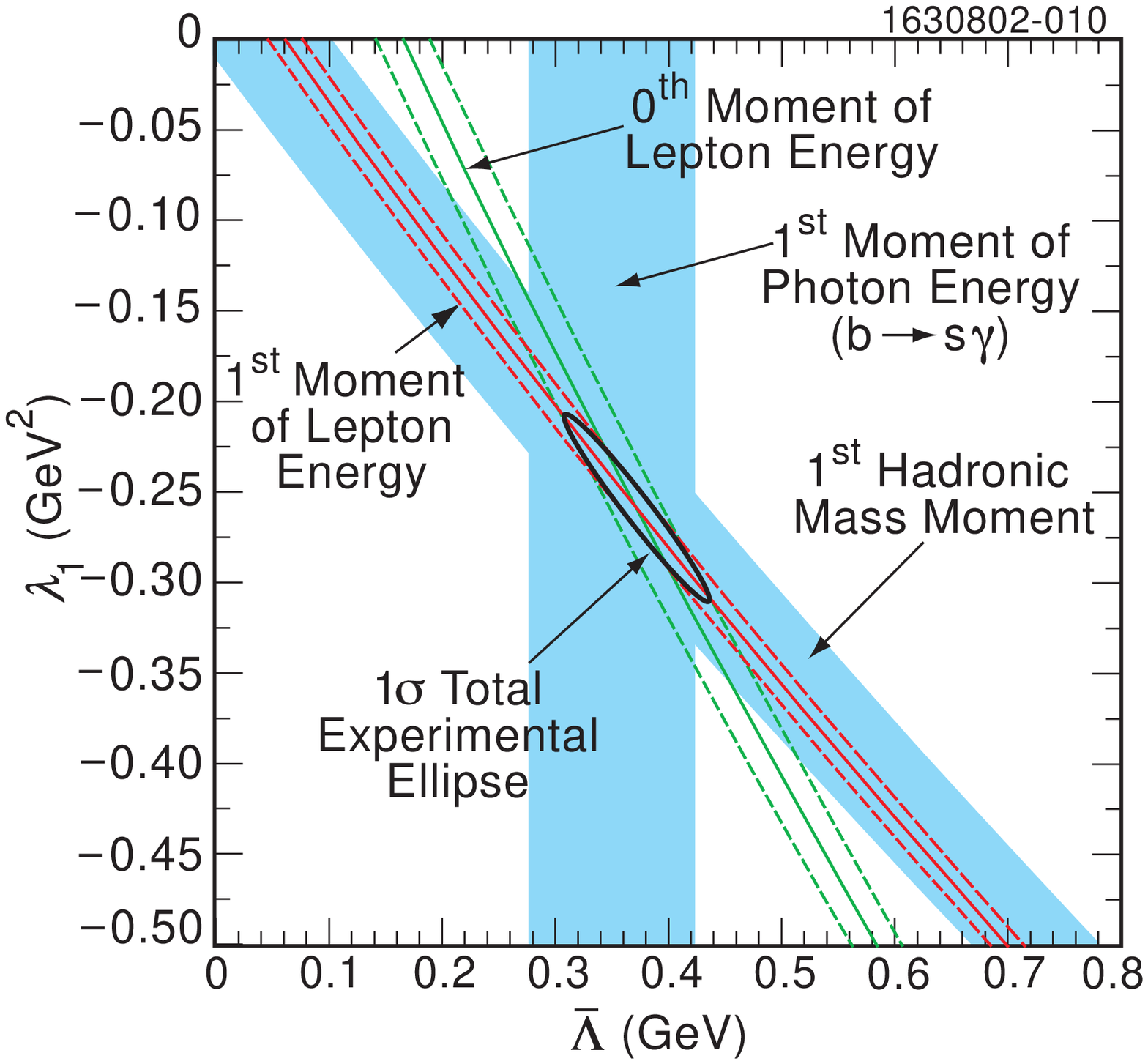,height=7.9cm}
    \end{minipage}
   }
 \end{minipage}
 \end{center}
 \vspace*{-0.8cm}
\end{figure}

An alternative approach to get \vcb is using measurements of the inclusive semileptonic  
decay rate \mbox{$B \to X_cl\bar{\nu}$}. Although there are fewer details of the hadronic final state to
be considered in inclusive measurements, non-perturbative effects do still affect 
the \vcb extraction from the measured rates. Again, HQET is used to control these effects.
Using an operator product expansion one can express moments of measured inclusive observables like lepton
energy and masses of hadronic final states in terms of HQET parameters $\bar{\Lambda},\, \lambda_1$ and
$\lambda_2$. These moments can be thought of as the mass difference between the \mbox{$b$-quark} and the $B$-meson 
$(\bar{\Lambda})$, the kinetic energy of the \mbox{$b$-quark} inside the $B$-meson $(\lambda_1)$ and the hyperfine interaction 
of the $b$-spin with the light degrees of freedom $(\lambda_2)$. By measuring moments of various inclusive 
distributions and extracting the HQET parameters one can reduce
the uncertainties on them and thereby the uncertainty in the extraction
of \vcb which can also be expressed in terms of HQET parameters. 
To extract the HQET parameters one can also make use of measurements of different quark-level processes which 
depend on the HQET parameters in similar ways, making the extraction less prone to systematic effects which 
are specific to semileptonic events.
The CLEO collaboration measures the first and the second moment of the photon energy spectrum of 
radiative $b \to s\gamma$ decays and combines it with a measurement of the mass moments of semileptonic
\btoc transitions to extract \vcb.
The photon energy spectrum from $b \to s\gamma$ decays is shown in Fig.~\ref{fig:bsgspectrum} \cite{bsgpaper}.\\
\\
The moments we extract from this are  
$\langle E_{\gamma} \rangle = (2.346 \pm 0.032 \pm 0.011) \gev$ and 
$\langle E^2_{\gamma} \rangle - \langle E_{\gamma} \rangle^2 = (0.0226 \pm 0.0066 \pm 0.0020) \gev^2$.
In a second analysis we measure the moments of the mass spectrum of the hadronic final state of 
$\bar{B} \to X_cl\nu$ decays, shown in Fig.~\ref{fig:massmoments}, using the hermeticity of 
the detector to reconstruct the neutrino. We get \cite{hadmoments} 
$\langle (M^2_x - \bar{M^2_D}) \rangle = (0.251 \pm 0.023 \pm 0.062) \, \gev^2$ and 
$\langle (M^2_x - \langle M^2_x \rangle ) \rangle = (0.576 \pm 0.048 \pm 0.163) \, \gev^4 $.
Combined with theoretical expressions for the first hadronic mass moment \cite{hqetmassmoments} these 
measurements provide a constraint on the HQET parameters $\bar{\Lambda}$ and $\lambda_1$ as 
shown in Fig.~\ref{fig:leptonmoments}.
The numerical values for the first and the second HQET parameter we get from this are 
$\bar{\Lambda} = (0.35 \pm 0.07 \pm 0.10) \gev $ and
$\lambda_1 = (-0.236 \pm 0.071 \pm 0.078) \gev^2 $.
Using these values, combined with a the theoretical expression for the semileptonic
decay width $\Gamma_{sl}$, and the measured value for the width,
$\Gamma_{sl} = (0.427 \pm 0.020) \times 10^{-10} \, \mathrm{MeV} \,$, which we get from
the total semileptonic branching ratio 
${\cal B}(B \to X_cl\nu) = (10.39 \pm 0.46)$ \cite{semiwidth}, and 
combined with the $B$-lifetime $\tau_{B^0} = (1.653 \pm 0.028)$ \cite{blifetime} we extract :\\ 
\vspace*{-0.4cm} \begin{center} $|V_{cb}| = (4.04 \pm 0.09 \pm 0.05 \pm 0.08) \times 10^{-2}$. \end{center}
Under the assumption of quark-hadron duality this represents a 3.2 \% measurement 
of \vcb.\\
In a third analysis we measure the inclusive lepton energy spectrum in semileptonic $B$ decays.
In Fig.~\ref{fig:leptonenergy} we show the measured spectrum which has $B\bar{B}$-backgrounds subtracted 
using tuned monte carlo simulations and $e^+e^- \to qq$-backgrounds subtracted using events 
recorded below the $\Upsilon (4S)$ resonance. The spectra have 
also been corrected  for efficiencies, detector resolutions, the motion of the $B$ in the lab 
frame and final state radiation.\\
\\
We compute the generalized moments :\\*[0.1cm]
\mbox{
$ R_0 = \frac{ \int_{1.7} \frac{d\Gamma}{dE_sl} dE_l } {\int_{1.5} \frac{d\Gamma}{dE_sl} dE_l  } 
      = 0.6187 \pm 0.0014 \pm 0.016 $ ;

$ R_1 = \frac{ \int_{1.5} E_l \frac{d\Gamma}{dE_sl} dE_l } {\int_{1.5} \frac{d\Gamma}{dE_sl} dE_l  } 
      = 1.7810 \pm  0.0007 \pm  0.0009 \gev$}\\*[0.1cm]
where the errors are statistical and systematic. Using HQET we extract from these moments  \\
$\bar{\Lambda} = (0.39 \pm 0.03 \pm 0.06 \pm 0.12) \gev$ and
$\lambda_1 = (-0.25 \pm 0.02 \pm 0.05 \pm 0.14) \gev^2$.\\
As shown in Fig.\ref{fig:leptonmoments} the values agree with the ones extracted from
the photon energy and the hadronic mass spectrum.
Finally, we can extract :\\
\vspace*{-0.5cm} \begin{center} $|V_{cb}| = (4.08 \pm 0.05 \pm 0.04 \pm 0.09) \times 10^{-2}$. \end{center} 
Again, the value is in good agreement with the value determined earlier. The agreement of the results 
is a nice confirmation of the self-consistency of the techniques used.
%
%
\section{Measurements of \vub}
Determinations of \vub are carried out in the same spirit as for \vcb via the measurement
of decay rates for processes with underlying quark-level transition \btou. 
Experimentally this is however more challenging since 
\btoc transitions present a hundred-fold larger background to \btou transitions.
As for \vcb one also uses semileptonic decays $B \to X_ul\nu$ to determine \vub since non-perturbative effects are 
much smaller and easier to control than in hadronic decays, 
although typically not to the same level as in \btoc transitions.
The large energy release and the fact that the final state particles are very light make it harder
to control these effects theoretically.
\subsection{\vub exclusive}
The exclusive measurement of \vub uses $B \to \pi l \nu$ and $B \to \rho l \nu$ events.
To unambiguously identify these events the neutrino is reconstructed from the missing 
energy and the missing momentum assuming that all other final state particles are detected.\\
\begin{figure}[h!]
\begin{center}
 \vspace*{-0.7cm}
 \begin{minipage}[c]{0.49\textwidth}
  \epsfig{figure=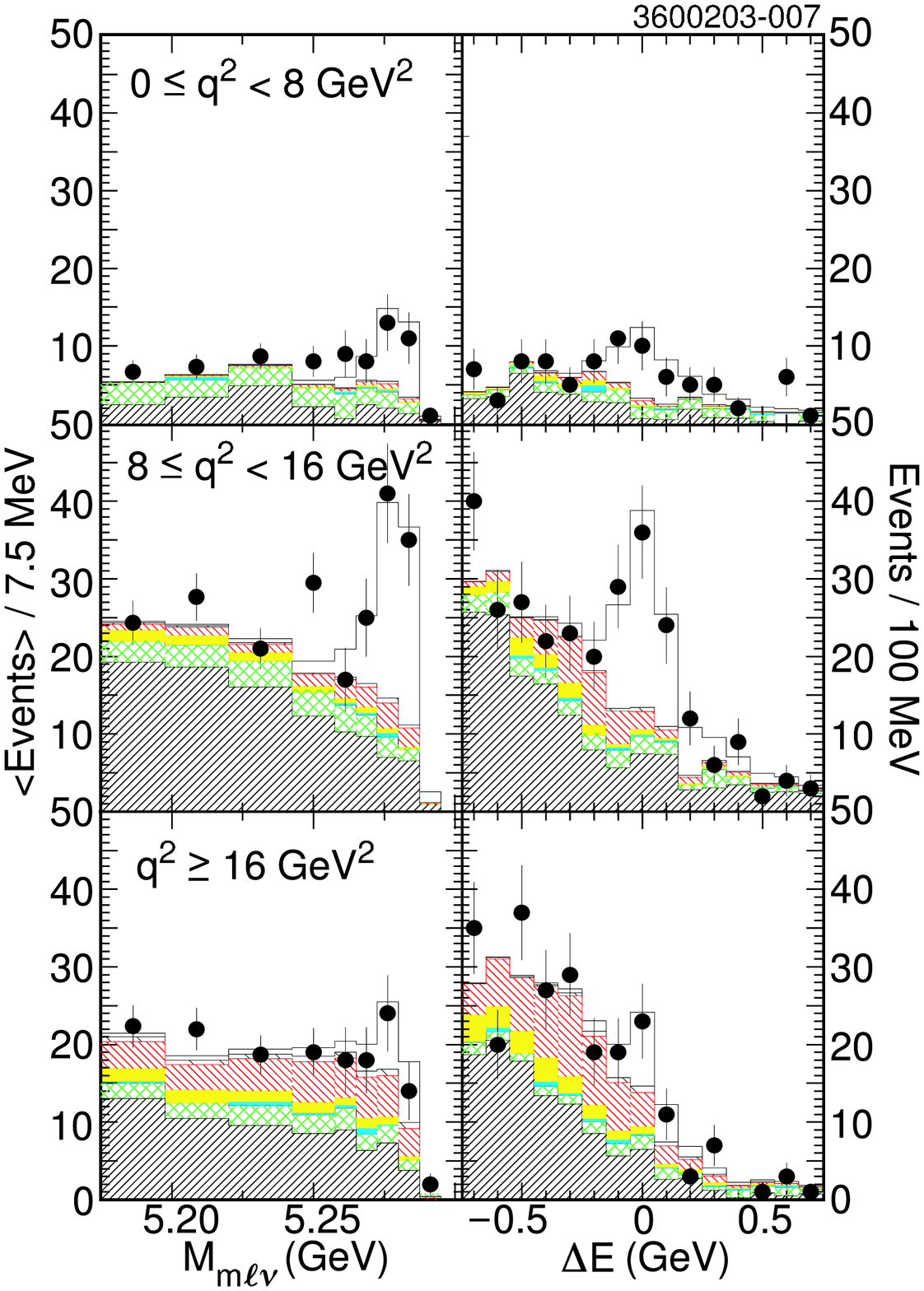,height=9.5cm}
  \vspace*{-0.5cm}
  \caption{Projections of $M_{ml\nu}$ and $\Delta E$ in three different bins of $q^2$ for the 
           combined $\pi^{\pm}$ and $\pi^0$ modes.
           \label{fig:vubexclusiveplots}}
 \end{minipage}
 \hfill
  {
  \begin{minipage}[c]{0.49\textwidth}
   \vspace*{0.2cm}
   \epsfig{figure=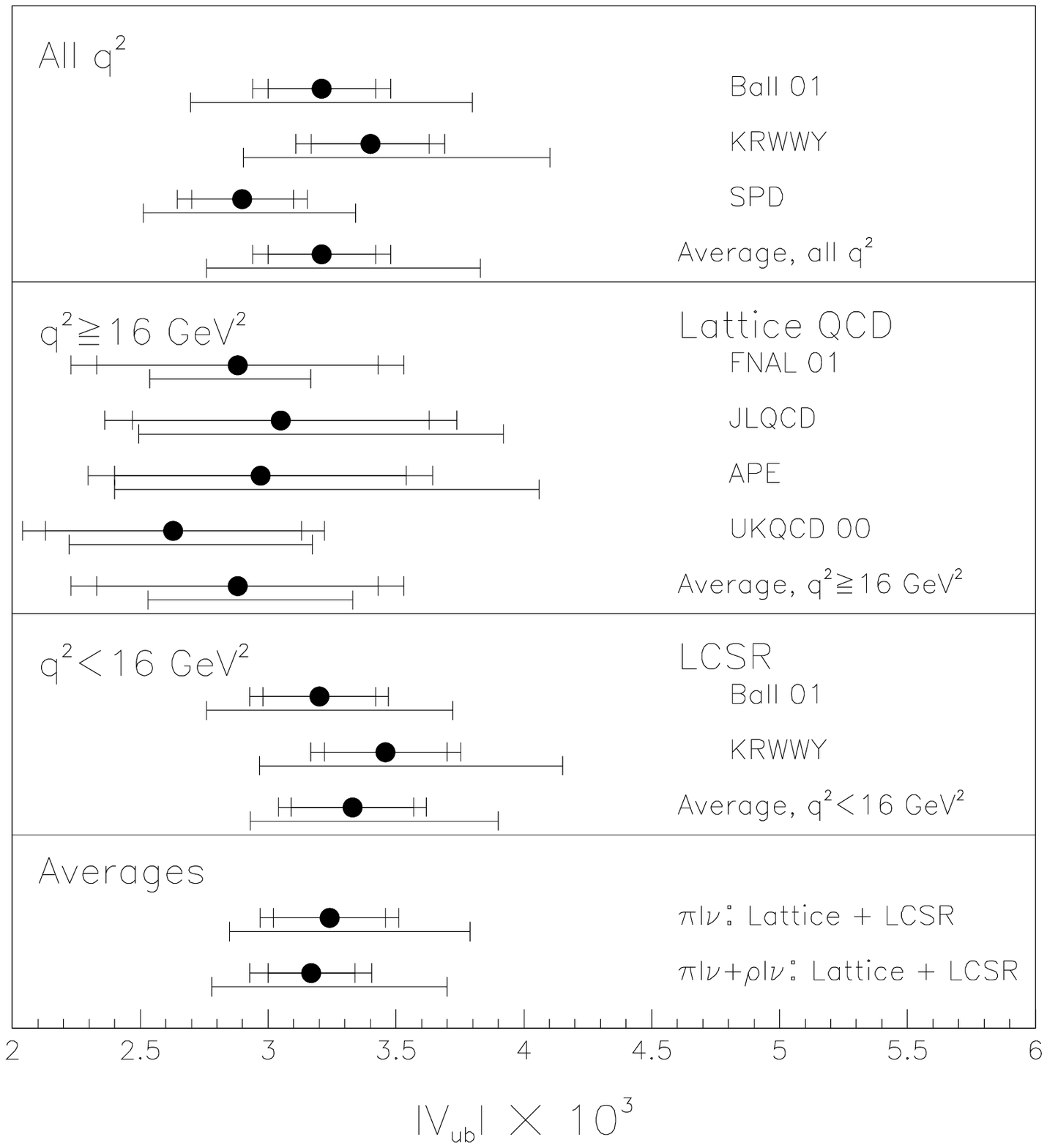,height=8.0cm} 
   \vspace*{-0.5cm}
   \caption{Extracted values for \vub from the exclusive semileptonic measurements of the $b \to ul\nu$ rate
           in different kinematic ranges using various theoretical approaches. 
           \label{fig:vubexclusive}}
  \end{minipage}
  }
 \end{center}
 \vspace*{-0.4cm}
\end{figure}\\
To better control non-perturbative effects the measurement was performed differentially in
the missing energy $\Delta E$, the meson mass $M_{ml\nu}$, the net charge in the event $\Delta Q$, 
the 2- (3)-$\pi$ meson mass ranges and $q^2$, the momentum transfer of the virtual $W$ in the event. 
The projections for $\Delta E$ and $M_{ml\nu}$ are shown in Fig.~\ref{fig:vubexclusiveplots}.
The branching ratios we extract from this are :\\
${\cal B}(B^0 \to \pi^- l^+ \nu ) = ( 1.33 \pm 0.18 \pm 0.11 \pm 0.01 \pm 0.07) \times 10^{-4}$,\\
${\cal B}(B^0 \to \rho^- l^+ \nu ) = ( 2.17 \pm 0.34 \pm 0.50 \pm 0.41 \pm 0.01) \times 10^{-4}$.\\
where the errors are  statistical, experimental systematic, systematic due to residual form-factor
 uncertainties in the signal, and systematic due to residual form factor uncertainties in cross feed
modes respectively.
To extract \vub from this we employ various theoretical calculations in different $q^2$ bins to minimize
model dependencies. For $q^2 < 16 \,\mathrm{ GeV^2}$ we use form-factor \mbox{shapes \cite{vubfromfactor}} and 
normalization results
from light-cone sum rules (LCSR) QCD calculations \cite{vubexcl} and for $q^2 \ge 16 \,\mathrm{ GeV^2}$   
we use lattice QCD (LQCD) \cite{vubexcl} studies. 
We average over the $\pi l \nu$ and 
 $\rho l \nu$ modes, and also combine, weighted by their errors, the LCSR and the LQCD extractions 
to get a best estimate of $|V_{ub}|$ \cite{vubexcl} :\\ 
\vspace*{-0.4cm} \begin{center} $|V_{ub}| = ( 3.17 \pm  0.17^{+0.16 +0.53}_{-0.17 -0.39} \pm 0.03) \times 10^{-3} $. \end{center}
The errors are statistical, experimental systematic, theoretical systematic from the LCSR and LQCD calculations and 
$\rho l\nu$ form-factor shape uncertainty, respectively.
In Fig.~\ref{fig:vubexclusiveplots} we show the individual results in a graphic representation,
demonstrating the very good agreement in different kinematic ranges and using different theoretical
calculations. 
%
%
\section{\vub inclusive}

\begin{figure}[h!]
\begin{center}
 \vspace{-0.8cm}
 \begin{minipage}[l]{0.47\textwidth} 
  \psfig{figure=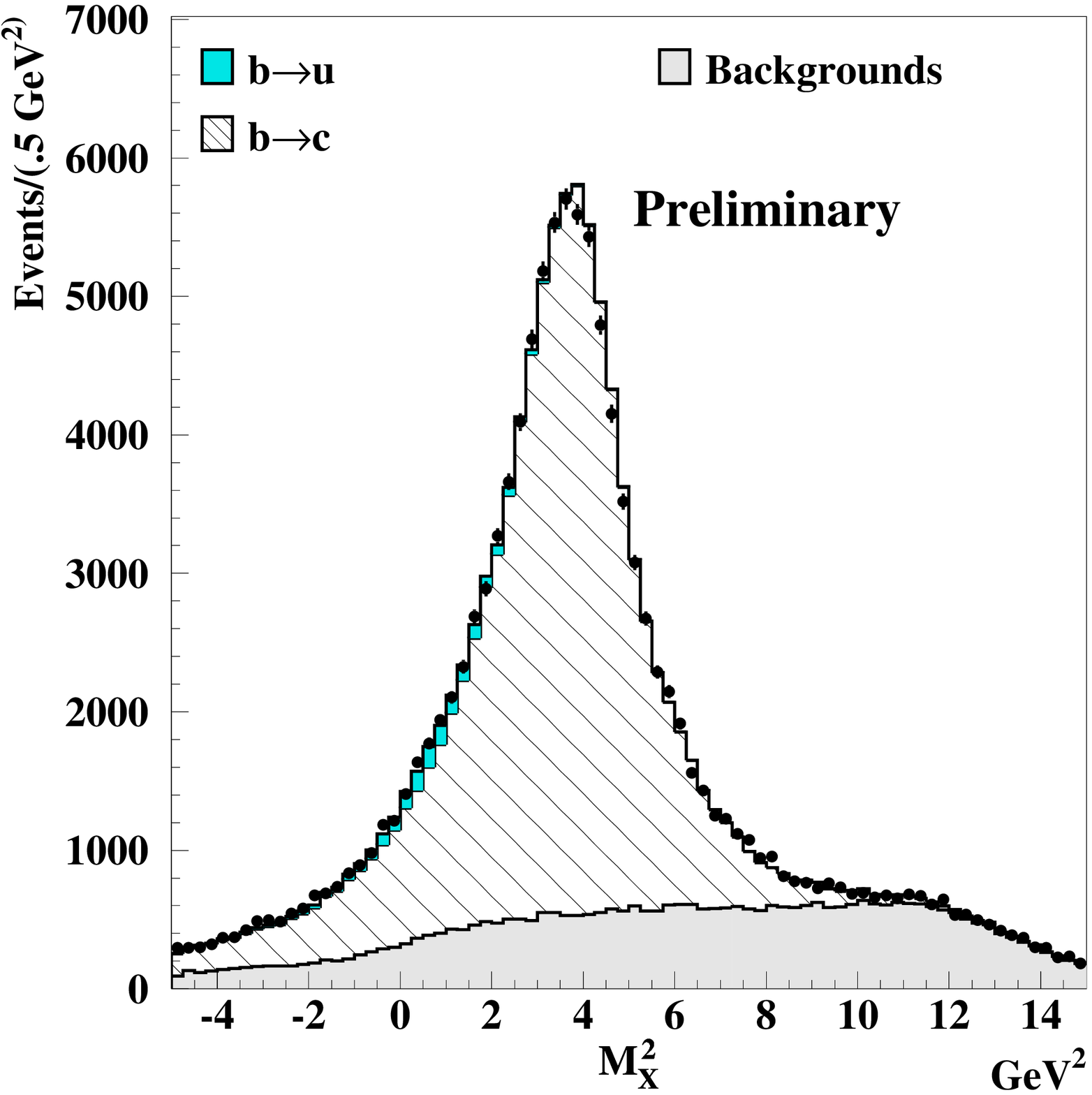,height=6.8cm}
  \vspace{-0.3cm}
  \caption{The $M^2_x$ distribution for the inclusive \vub analysis. Note that
           the \btou component is only a tiny part of the entire sample, visible 
           on the left shoulder of the peak in the $M^2_x$ distribution which is mainly
           from \btoc transitions and backgrounds. 
           \label{fig:vubmxinclusive}}
 \end{minipage}
  {
  \begin{minipage}[r]{0.47\textwidth}
   \psfig{figure=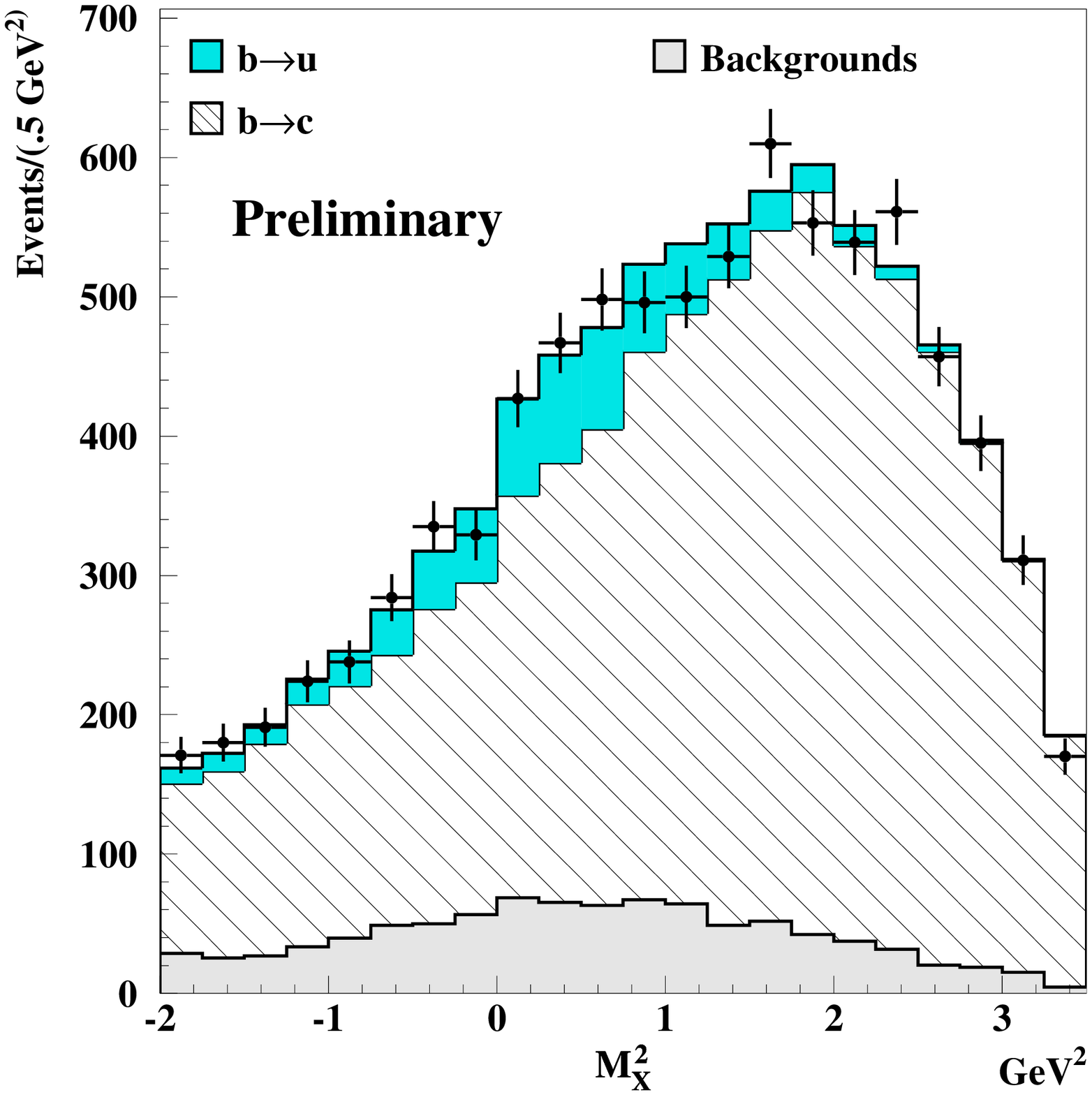,height=6.8cm}
   \vspace{-0.3cm}
   \caption{By selecting the region most sensitive to \vub in all three dimensions
           the \btou component becomes more pronounced. Note that the shape of the distribution
           is markedly different for \btou and \btoc transitions.
           \label{fig:vubmxinclusivezoom}}
  \end{minipage}
  }
 \end{center}
 \vspace{-0.3cm}
\end{figure}

The inclusive measurement of \vub takes advantage of the fact that the $u$-quark is lighter than the 
$c$-quark so that more kinetic energy is released in \btou transitions
than in \btoc transitions. Thus the momentum spectrum of final state particles, in particular the lepton momentum spectrum in semileptonic decays, is harder in \btou transitions than it is in \btoc transitions.
The relative yield of leptons from \btou transitions will therefore be enhanced towards the endpoint of the spectrum.
The challenge is to correctly model the shape of the momentum spectrum for both the \btou and \btoc 
transitions and subtract the latter from the total lepton yield to get the \btou contribution. 
One technique employed by CLEO makes use of the fact that smearing effects which change the shape of the lepton spectrum in \btou transitions when going from the parton to the hadron level are very similar to the smearing
of the photon energy spectrum in $b \to s\gamma$ transitions. Thus, we measure this photon energy spectrum
and extract a shape function which parameterizes this smearing. This shape function is then used to
model the shape of the lepton energy spectrum and separate out the contribution from \btou transitions \cite{bsgtechnique}.\\
\begin{figure}[h!]
\begin{center}
 \begin{minipage}[c]{0.58\textwidth}
  \vspace*{-0.7cm}
  \psfig{figure=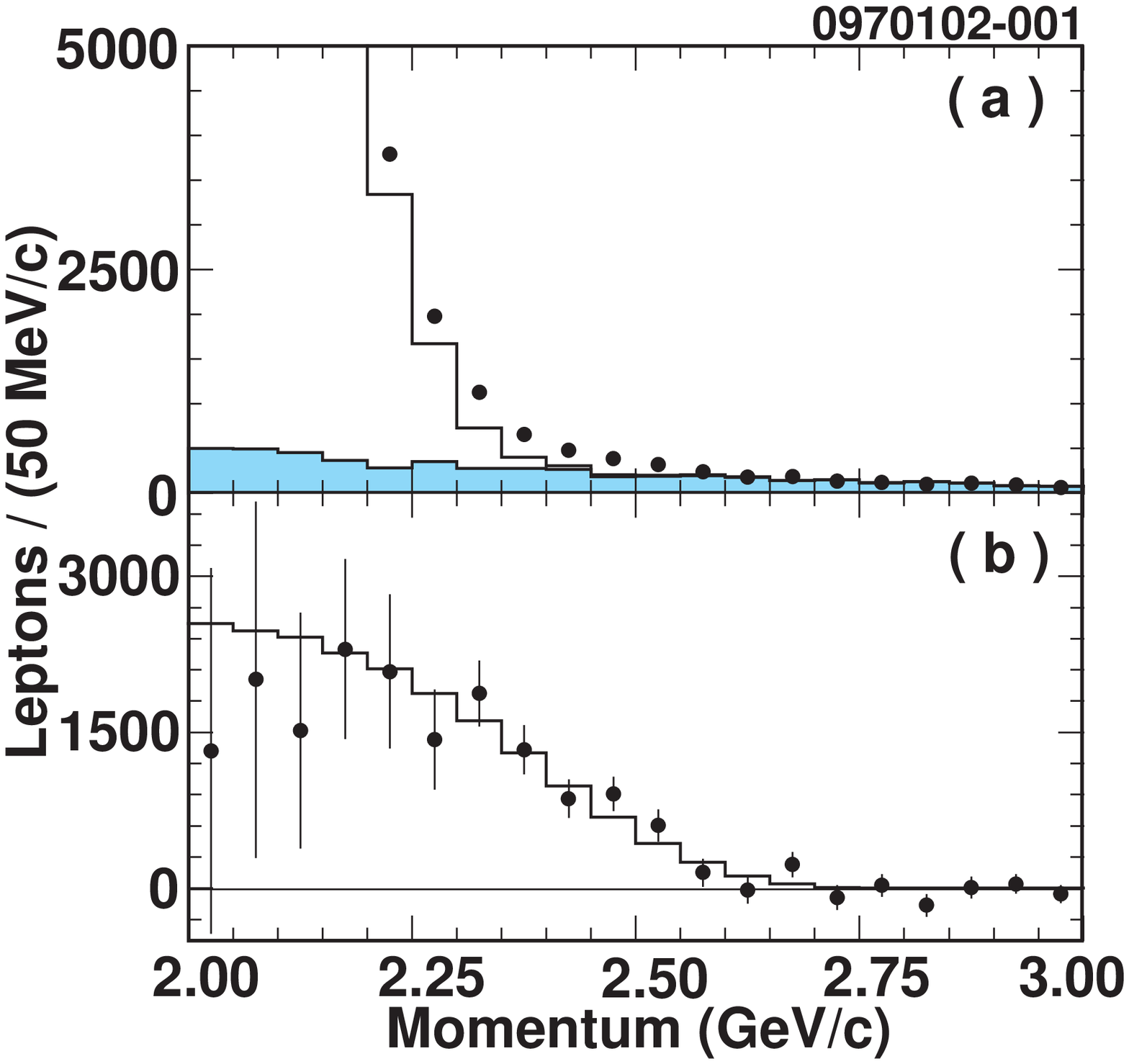,height=6.5cm}
  \vspace*{-0.5cm}
  \caption{The lepton energy spectrum at the endpoint for semileptonic $B$ decays. The top plot shows the
           CLEO data (points) compared to \btoc monte carlo and backgrounds (both histogram). The difference (points
           in the bottom plot) is due to the \btou contribution (solid line in the bottom plot).
  \label{fig:leptonendpoint}}
  
 \end{minipage}
 \hfill
  {
  \begin{minipage}[l]{0.40\textwidth}
  \hspace*{-1.0cm}
  \epsfig{figure=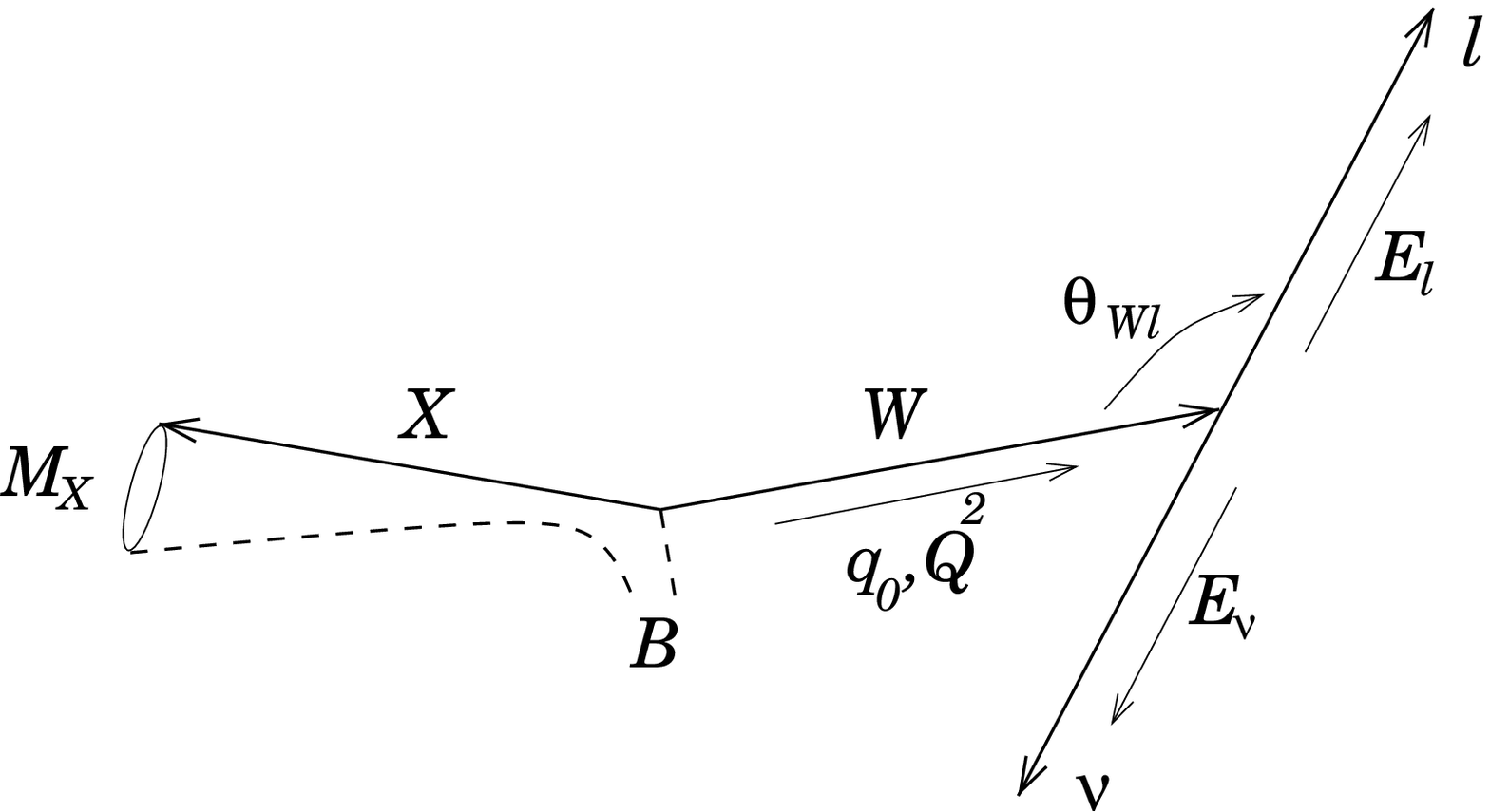,height=4.0cm}
  \caption{The diagram illustrates the kinematic variables $\mathrm{cos} \Theta_{Wl}$, $Q^2$ and $M_x$ which
   are used in the three-dimensional fit to measure the \btou rate from semileptonic $B$ decays.
  \label{fig:incldiagram}}
   
  \end{minipage}
  }
 \end{center}
 \vspace*{-0.5cm}
\end{figure}\\
The value of \vub we extract from the measured rate is \cite{vubincllep} :\\      
\begin{center} \vspace*{-0.4cm} $|V_{ub}| = ( 4.08 \pm  0.34 \pm 0.44  \pm 0.16 \pm 0.24) \times 10^{-3}$. \end{center}
The first two errors are experimental and the second two errors are from the shape function technique and
the \vub extraction from the measured \btou rate.\\ 
The lepton energy is not the only kinematic variable in semileptonic decays which can be used to separate
\btou from \btoc components. The distributions for mass of the hadronic system $M_x$, the angle between the 
virtual $W$ and the lepton, $\mathrm{ cos}(\Theta_{Wl})$, and the momentum transfer squared to the $W$, $Q^2$, 
also show distinct differences in shape between \btou and \btoc transitions. 
To fully exploit this information CLEO uses a fully inclusive three-dimensional fitting technique to 
measure a rate for \btou and extract \vub.
In Fig.~\ref{fig:vubmxinclusive} we show the mass spectrum of the inclusive semileptonic sample which is
used in the fit. In Fig.~\ref{fig:vubmxinclusivezoom} we show a projection in $M_x$, just for illustration purpose, 
for a subsample restricted in all three dimensions of the phase space to the area most sensitive to the \btou component. The contribution from \btou, barely
visible in Fig.:~\ref{fig:vubmxinclusive}, becomes much more pronounced. 
To better control systematic effects we not only fit for the \btou component but also include $B \to Dl\nu$,
$B \to D^{\star}l\nu$, $B \to D^{\star \star}l\nu$ and non-resonant $B \to X_c l\nu$ components separately. 
The backgrounds, $b \to c \to Xl\nu$ decays, $e^+e^- \to qq$ events with a lepton, or fake leptons, from
non-semileptonic $B$ decays, are treated as separate background components.\\
\\
To get to the total rate for $b \to ul\nu $ we use HQET to calculate the ratio between the portion 
of the three-dimensional phase space we cover in this analysis and the total phase space, 
using the formula
$\Delta B_{region} = {\cal B} (B \to X_u l \nu)^{model} \times f^{HQET}_{region}$.
From this we extract \vub using the relation \\
$ |V_{ub}| = [3.07 \pm 0.12 \times 10^{-3}]
 \left[
  \frac{\Delta B_{region}} {0.001 \times f^{HQET}_{region}} \frac{1.6 ps}{\times \tau_B}
 \right]^{1/2}$
as determined in \cite{bauerinclvub} which also pioneers the application of the multidimensional cuts
used here. Alternative calculations of the conversion factor yield 
very similar results \cite{neubertinclvub}.  
By optimizing the region used to extract \vub we can minimize the theoretical uncertainties
and get the preliminary result \cite{3dimvub} :\\
\begin{center} \vspace*{-0.4cm} $V_{ub} = ( 4.05 \pm 0.18 \pm 0.58 \pm 0.25 \pm 0.21 \pm 0.56) \times 10^{-3} $ \hspace*{0.2cm} (Preliminary). \end{center}
The errors are statistical, detector systematics, $B \to X_cl\nu$ model dependence, $B \to X_ul\nu$
model dependence and theoretical uncertainties respectively. The result is in agreement with 
the lepton endpoint analysis from CLEO presented above. There is a slight correlation with that 
analysis and the errors may be correlated as well. Studies of these correlations, further studies on  
the optimal region and extraction of moments are in progress. 
%
%
\section{Summary}

CLEO uses a broad range of experimental techniques to measure \vub and \vcb in inclusive and 
exclusive semileptonic $B$-decays. These measurements, which impose important constraints on the 
Unitarity Triangle, are always systematically limited. By using different techniques we 
gain deeper insight into experimental and theoretical details of how to best determine
these important quantities. To extract the fundamental parameters \vub and \vcb from measured
rates we depend on HQET and lattice QCD. The future CLEO-c program will help to solidify 
the experimental basis of these theoretical approaches which are essential to progress in the field 
of heavy flavor physics in general.  

\section*{Acknowledgments}
We gratefully acknowledge the contributions of the CESR staff for providing the luminosity and
and the National Science Foundation and the U.S. Department of Energy for supporting this research.
The author wishes to thank the organizers of the 'Rencontres de Moriond' for a very exciting and inspiring event.

\end{document}